\documentclass{iau} 
\usepackage{graphicx}
\def\KMS{kms$^{-1}$}

\title[PDR Emission from the Arched-Filaments and Nearby Positions.] 
{PDR Emission from the Arched-Filaments and Nearby Positions\\}

\author[P. Garc\'ia, M. R{\"o}llig, N. Abel, M. Steinke, M. Burton, \& R. Blackwell]   
{Pablo Garc\'ia$^1$,
Markus R{\"o}llig$^2$,
Nicholas Abel$^3$,
Martin Steinke$^2$,
Michael Burton$^4$,
\and Rebecca Blackwell$^5$}

\affiliation{$^1$Instituto de Radioastronomía Milimétrica IRAM\\ Av. Divina Pastora 7, 18012 Granada, Spain\\ email: {\tt pgarcia@iram.es}\\[\affilskip]
$^2$I. Physikalisiches Institut, University of Cologne, Germany \\[\affilskip]
$^3$University of Cincinnati, USA \\[\affilskip]
$^4$Armagh Observatory and Planetarium, Northern Ireland\\[\affilskip]
$^5$School of Physical Sciences, University of Adelaide, Australia}

\pubyear{2016}
\volume{322}  
\jname{The Multi-Messenger Astrophysics of the Galactic}
\editors{S. Longmore, G. Bicknell, \& R. Crocker, eds.}
\begin{document}

\maketitle

\begin{abstract}
We investigate the physical conditions of the gas, atomic and molecular, in the filaments in the context of Photo-Dissociation Regions (PDRs) using the KOSMA-PDR mode of clumpy clouds. We also compare the [CII] vs. [NII] integrated intensity predictions in Abel et al. 2005 for HII regions and adjacent PDRs in the Galactic disk, and check for their applicability under the extreme physical conditions present in the GC. Our preliminary results show that observed integrated intensities are well reproduced by the PDR model. The gas is exposed to a relatively low Far-UV field between 10$^{2}$ - 10$^{3}$ Draine fields. The total volume hydrogen density is well constrained between 10$^{4}$ - 10$^{5}$ cm$^{-3}$. The hydrogen ionization rate due to cosmic-rays varies between 10$^{-15}$ and 4$\times$10$^{-15}$ s$^{-1}$, with the highest value $\sim$ 10$^{-14}$ s$^{-1}$ found towards G0.07+0.04. Our results show that the line-of-sight contribution to the total distance of the filaments to the Arches Cluster is not negligible. The spatial distribution of the [CII]/[NII] ratio shows that the integrated intensity ratios are fairly homogeneously distributed for values below 10 in energy units. Calculations including variation on the [C/N] abundance ratio show that tight constraints on this ratio are needed to reproduce the observations.

\end{abstract}

\firstsection 
\section{Introduction}
In the Galactic Center (GC), the Arched-Filaments are thought to originate from the interaction of the gas with the radiation field of the massive Arches Cluster [\cite[Simpson et al.(2007)]{simpson2007}]. Using the sub-mm data in \cite[Garc\'ia et al.(2016)]{garcia2016}, and complementary millimeter data from the Mopra archive [\cite[Jones et al.(2012)]{jones2012}], and unpublished CO(1-0) and 13CO(1-0) rotational lines from the ongoing Mopra CMZ Survey [R. Backwell et al., in prep.], we investigate the physical conditions of the gas, both atomic and molecular, in the Arched-Filaments region in the context of Photo-Dissociation Regions (PDRs) using the KOSMA-PDR model of clumpy clouds [\cite[Stoerzer et al.(1996)]{stoerzer1996}, \cite[Cubick et al.(2008)]{cubick2008}, and \cite[R{\"o}llig et al.(2013)]{rollig2013}]. We also compare the [CII] vs. [NII] integrated intensity predictions in \cite[Abel et al.(2005)]{abel2005} for HII regions and adjacent PDRs in the Galactic disk, and check for their applicability under the extreme physical conditions present in the GC.

\section{Overview}

We have selected seven positions within the Arched-Filaments following the peaks of 20 cm continuum emission in \cite[Yusef-Zadeh \& Morris(1987)]{yusef1987} and two positions close to them tracing high density gas in \cite[Serabyn \& Guesten(1987)]{serabyn1987}. From the KOSMA-PDR model of clumpy clouds, the following quantities are obtained from the fit to the observed integrated intensities: hydrogen nucleus number density n(H)$_{s,tot}$; total clump mass within the beam M$_{cl}$; FUV field in Draine units $\chi$; hydrogen ionization rate due to cosmic-rays $\eta$(H); clump ensemble mass lower limit M$_{low}$; and clump ensemble mass upper limit M$_{upper}$. The model intensities can reproduce the observed intensities extremely well in almost all cases within error bars, except for the CO(1-0) transition where the model systematically underestimates observations by a factor of $\sim$ 2 - 3. Calculations for the contribution of HII regions to the observed [CII] intensities as traced by [NII] emission for Galactic plane sources have been presented in \cite[Abel et al.(2005)]{abel2005}. When compared with the observed intensities in and around the Arched-Filaments location, their predictions for solar and under-solar metallicities do not match the data. The observations show that the great majority of the integrated intensity [CII]/[NII] ratios (calculated between $-$70 \KMS\textrm{ }and $+$20 \KMS) are fairly homogeneously distributed over the map for values below 10. Calculations including variations in the [C/N] abundance ratio and stellar clusters SEDs can explain the observed scatter in the [CII] vs. [NII] plot and suggest that, in order to use the [NII] intensity to disentangle the [CII] component arising from H II regions, tight constraints are needed on the [C/N] abundance ratio, significantly tighter than previous abundance measurements have discerned.

\section{Implications}

Our preliminary results show that the gas is exposed to a relatively low $\chi$, between 10$^{2}$ - 10$^{3}$ Draine fields, with n(H)$_{s,tot}$ well constrained between 10$^{4}$ - 10$^{5}$ cm$^{-3}$. The $\eta$(H) values vary between 10$^{-15}$ and 4$\times$10$^{-15}$ s$^{-1}$, with the highest value $\sim$ 10$^{-14}$ s$^{-1}$ found towards G0.07+0.04. The model results (specially for n(H)$_{s,tot}$ and $\chi$), for all positions as a function of projected distance from the Arches Cluster, show a large dispersion suggesting that the true distance to the cluster must have a significant line of sight component. Combining all these results puts constraints on the 3-D spatial distribution, 3-D density structure, and hydrogen ionization rate due to cosmic-rays in the Arched-Filaments. We discuss this further in an upcoming publication.

\end{document}